\documentclass[a4paper,12pt]{scrartcl}
\usepackage{enumitem}
\usepackage{amssymb}
\usepackage{amsmath}
\usepackage{subcaption}
\usepackage[english]{babel}
\usepackage{makecell}
\usepackage{color}
\usepackage{booktabs}
\usepackage[cp1251]{inputenc}
\usepackage{graphicx}
\usepackage[unicode, bookmarksopen=true]{hyperref}
\usepackage{wasysym}
\usepackage{xcolor}
\usepackage[symbol*]{footmisc}
\usepackage{pifont}
\usepackage{listings}
\usepackage{setspace}

\definecolor{Code}{rgb}{0,0,0}
\definecolor{Decorators}{rgb}{0.5,0.5,0.5}
\definecolor{Numbers}{rgb}{0.5,0,0}
\definecolor{MatchingBrackets}{rgb}{0.25,0.5,0.5}
\definecolor{Keywords}{rgb}{0,0,1}
\definecolor{self}{rgb}{0,0,0}
\definecolor{Strings}{rgb}{0,0.63,0}
\definecolor{Comments}{rgb}{0,0.63,1}
\definecolor{Backquotes}{rgb}{0,0,0}
\definecolor{Classname}{rgb}{0,0,0}
\definecolor{FunctionName}{rgb}{0,0,0}
\definecolor{Operators}{rgb}{0,0,0}
\definecolor{Background}{rgb}{0.98,0.98,0.98}

\usepackage{indentfirst}
\begin{document}
\title{Time Series Analysis: \\ yesterday, today, tomorrow}
\author{Igor Mackarov \\ \href{mailto:Mackarov@gmx.net}{{\small Mackarov@gmx.net}}}
\date{}
\maketitle
\begin{abstract}
	Forecasts of various processes have always been a sophisticated problem for statistics and data science. Over the past decades the solution procedures were updated by deep learning and kernel methods. According to many specialists, these approaches are much more precise, stable, and suitable compared to the classical statistical linear time series methods. Here we investigate how true this point of view is.


\end{abstract}

\section{Introduction}
Developments in deep learning and other non-linear methods of data science have significantly influenced, in particular, such its domain as time series analysis aimed at  revealing hidden dependencies and forecast of the future. Investigate these new methods in comparison with the traditional statistical methods in action.

\paragraph{The data} to be considered are in a CSV table available {\href{https://data.world/hhaveliw/airplane-crashes-1908-2009}{\textit{\textcolor{red}{here}}}. Each the table line reports in detail about the flight that ended in crash (time, place of departure; time, place of crash, aircraft brand, company, detailed description of the circumstances, etc.). Supposedly, collected are all the tragic flights from 1908 to 2009. Based on this information, we constructed  time series for test forecasts of the number of disasters within certain time intervals.

\paragraph{The code} implementing all the below described tests and preliminary diagnostics is  {\href{https://github.com/AndreoBochelli/TS_Analysis}{\textit{\textcolor{red}{here}}}. It contains means for forming a time series with an \textit{arbitrary} time step (class \textit{{crashes\_utils.Crashes\_prepare\_timedelta}}). Hopefully, all the information needed for work with it is in README.md.

\paragraph{Sampling rate} should be a reasonable trade-off between preserved details of the original data and the time series regularity. Here we will present results for intervals 6, 10, and 12 months with portions of test data 20\%, 20\%, and 10\%, respectively (Fig. \ref{fig:ts} ).

\begin{figure}[!h]
	\centering
	\begin{subfigure}[]{0.32\textwidth}
		\includegraphics[width=1\textwidth]{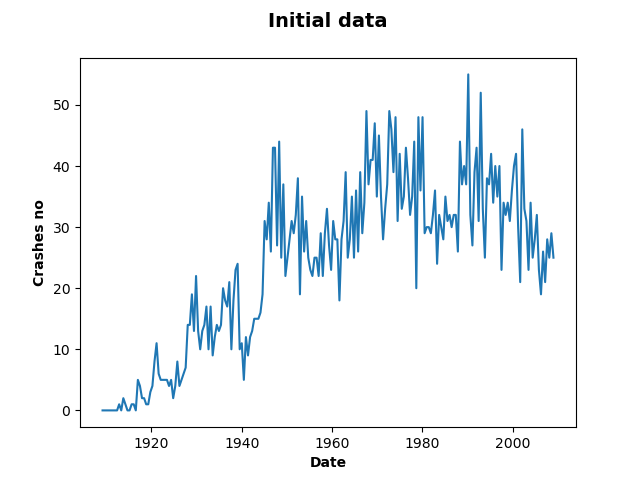}
		\caption{6 months interval}
			\label{fig:ts_12}
	\end{subfigure}
	\begin{subfigure}[]{0.32\textwidth}
		\includegraphics[width=1\textwidth]{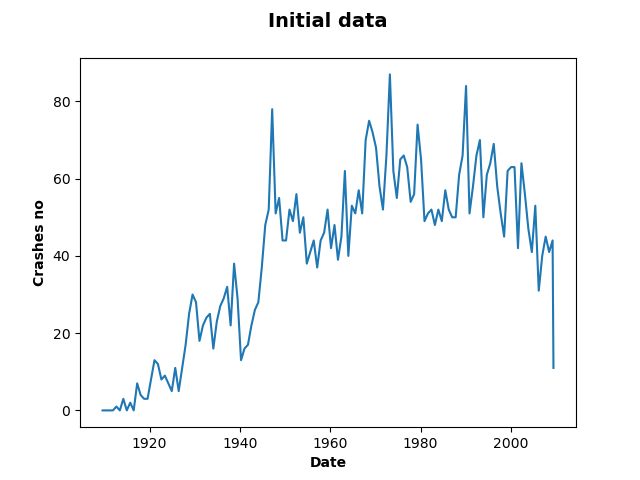}
		\caption{10 months interval}
	\end{subfigure}
	\begin{subfigure}[]{0.32\textwidth}
		\includegraphics[width=1\textwidth]{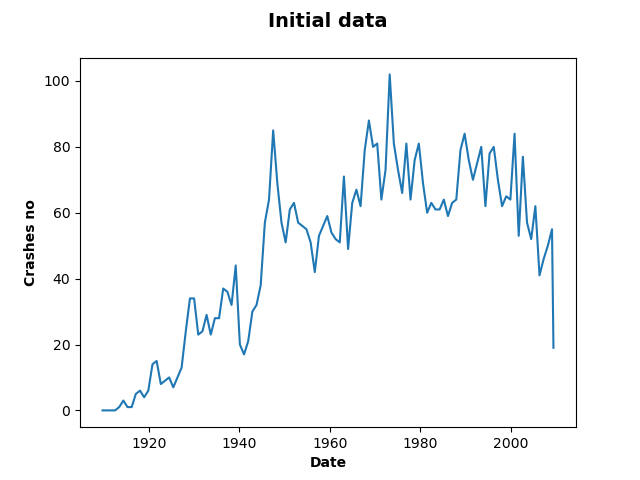}
		\caption{12 months interval}
	\end{subfigure}
	\caption{Time series aggregated over various time intervals}
	\label{fig:ts}
\end{figure}

\section{Traditional statistical models for time series}
\subsection{Requirements for  time series stationarity, its study, ways to achieve it}

According to \cite{Vishwas, Nielsen}, traditional statistical models require stationarity of the \textit{weak type}, that is, the series mean, variance, and covariance are constant over the time.

\subsubsection{Methods for studying the original time series stationarity}
\paragraph{Time series decomposition into components} (\textit{\textbf{the Code}}, module \textit{crashes\_ARIMA})
There exist visual decomposition methods that allow to visually control how stationary a series is. For example:
\begin{lstlisting}[language=Python]
def decompose(ts, decomposition_period):
    from statsmodels.tsa.seasonal import seasonal_decompose
    decomposition = seasonal_decompose(ts, model='additive',
                                       period=decomposition_period)
    trend = decomposition.trend
    seasonal = decomposition.seasonal
    residual = decomposition.resid
    # ...
\end{lstlisting}
Fig. \ref{fig:decomp_init}  shows what the components of the original series look like. The components are:

 \begin{itemize}
 	\item[\ding{42}] \textit{Trend} --- change of the time series mean values over some constant moving interval,
 	\item[\ding{42}]  \textit{Seasonality} ---variations on the scale of specific time intervals, the seasons,
 	\item[\ding{42}]  \textit{Residuals} --- what has rested after eliminating the two main factors of non-stationarity.
 \end{itemize}

\begin{figure}[!b]
	\centering
		\includegraphics[width=1\textwidth]{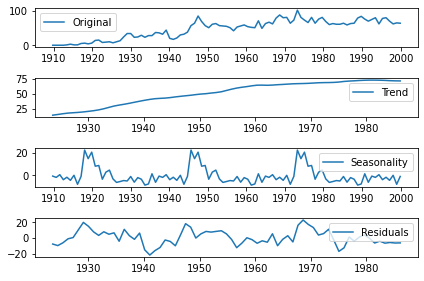}
	\caption{Decomposition of the original time series in Fig. \ref{fig:ts_12}}
	\label{fig:decomp_init}
\end{figure}

We see that the trend shows a significant tendency to increase, so the series is hardly stationary. This is \textit{formally} confirmed by...

\paragraph{Augmented Dickey-Fuller test}
(\textit{\textbf{the Code}}, module \textit{crashes\_ARIMA})
\begin{lstlisting}[language=Python]
from statsmodels.tsa.stattools import adfuller
#...
def Augmented_Dickey_Fuller_Test_func(series, column_name):
    print(f'Results of Dickey-Fuller Test for column: {"Crashes No"}')
    dftest = adfuller(series, autolag='AIC')
    dfoutput = pd.Series(dftest[0:4],
                         index=['Test Statistic', 'p-value', 'No Lags Used', 'Number of Observations Used'])
    for key, value in dftest[4].items():
        dfoutput['Critical Value (%s)' % key] = value
    print(dfoutput)
    if dftest[1] <= 0.05:
        print("Conclusion:====>")
        print("Reject the null hypothesis")
        print("Data is stationary")
    else:
        print("Conclusion:====>")
        print("Fail to reject the null hypothesis")
        print("Data is non-stationary")
\end{lstlisting}
 {\scriptsize
\textbf{{
 \begin{tabular}{@{} l *4c @{}}
\multicolumn{1}{l}{Results of Dickey-Fuller Test:}    &     \\
Test Statistic  & -1.807001 \\
p-value  & 0.376992  \\
\#Lags Used  & 2.000000  \\
Number of Observations Used  & 96.000000  \\
Critical Value (1\%)   &         -3.500379 \\
Critical Value (5\%)    &        -2.892152 \\
Critical Value (10\%)   &        -2.583100 \\
dtype: float64   \\
Conclusion:  ====> \\
Fail to reject the null hypothesis \\
Data is non-stationary
\end{tabular}}}}

\vspace{15pt}
\noindent \textit{p-value} is really big, so for successful forecasting it is necessary to apply...

\subsubsection{Tools for achieving stationarity}
\label{stable}

\paragraph{Transformation of the series variable} may be useful, for example, when examining the number of illnesses or deaths as a pandemic grows exponentially. A logarithmic transformation will `take out of the brackets' the avalanche-like increase in the number of diseases. It will thus allow to focus on volatility depending on the quality of medicine, the government activities, the level of public health, etc.

As to our data,  it resemble the $\sin$ function. So an idea has come to apply the $\arcsin$ transformation. As a result, a series was obtained  (Fig.~ \ref{arcus}), consisting of two more or less linearly stationary parts, connected by a steeply increasing isthmus. The isthmus origin is clear: near the $\sin$  maximum its derivative is close to 0, whereas the derivative of $\arcsin$ is big.

Despite this detail, the $\arcsin$--transformation turned out quite practical (see section~\ref{sec:kernel_results}).

\begin{figure}[!h]
	\centering
	\includegraphics[width=0.8\textwidth]{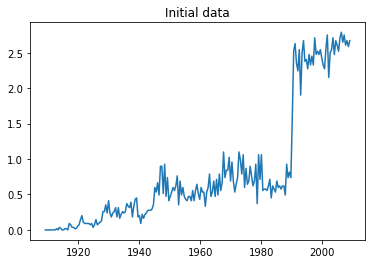}\\
	\caption{$\arcsin$-transformed data}
	\label{arcus}
\end{figure}

\paragraph{Differencing}
\label{sec:diff}
often proves practical and effective. In \textit{\textbf{the Code}}, for this and reverse operations functions \textit{diff, undiff} of class \textit{ARMA} were used  (file \textit{crashes\_utils.py}).
Look at Figure \ref{fig:decomp_diff}. This is how the decomposition of the differenced series looks. Comparing it to Fig.~\ref{fig:decomp_init}, we see that the trend now no more  has a distinct tendency. Is the stationarity reached? It certainly is. The Dickey-Fuller test confirms this:

 {\scriptsize
	\textbf{{
			\begin{tabular}{@{} l *4c @{}}
				\multicolumn{1}{l}{Results of Dickey-Fuller Test:}    &     \\
				Test Statistic  & -9.410135e+001 \\
				p-value  &  5.865458e-16  \\
				\#Lags Used  & 1.000000  \\
				Number of Observations Used  & 1.070000e+02  \\
				Critical Value (1\%)   &         -3.492996 \\
				Critical Value (5\%)    &        -2.888955 \\
				Critical Value (10\%)   &         -2.581393 \\
				dtype: float64   \\
				Conclusion:  ====> \\
				Reject the null hypothesis \\
				Data is stationary
\end{tabular}}}}

\begin{figure}[!h]
	\centering
	\includegraphics[width=1\textwidth]{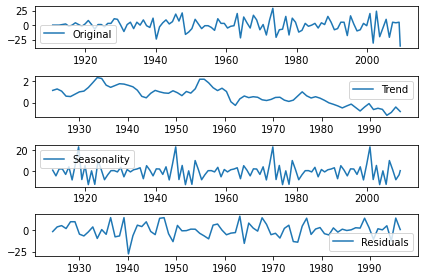}
	\caption{Decomposition of the differenced series of Fig. \ref{fig:ts_12}}
	\label{fig:decomp_diff}
\end{figure}

\paragraph{Moving average} \label{par:ma} consists in replacing the current value of a series with the average of several previous values. The \textit{exponentially weighted} one  gives higher weights to recent values and lower weights to more distant ones.

In our case, this technique proved not as good as differencing: a significant seasonality remained. Another inconvenience: it is impossible to restore the forecast done to the original terms.

\subsection{Autoregressive (AR) and Moving Average (MA) models}

\textbf{AR}:

\begin{equation}
	{X}_t=\phi_0+\phi_1 X_{t-1}+\phi_2 X_{t-2}+...+\phi_q X_{t-q}.
	\label{eq:AR}
\end{equation}
Here all $\phi$ are constants. Thus, the currently predicted value $X_t$, according to the model, entirely depends on a certain number of nearby historical values.

\textbf{MA}:
\begin{equation}
	{X}_t=\mu+\epsilon_t+\theta_1 \epsilon_{t-1}+\theta_2 \epsilon_{t-2}+...+\theta_q \epsilon_{t-q}
	\label{eq:MA}
\end{equation}
with $\mu$ --- \textit{average} value of the entire series, $\epsilon$ --- \textit{errors} introduced into the process at correspondent moments, $\theta$ --- constants. Therefore, from the point of view of MA, the variability of the time function is entirely due to the influence of random errors (life is unexpected, but sometimes predictable). Since the average of an error over the entire time range is 0, the intercept term in (\ref{eq:MA}) cannot be anything other than the average of the series over the entire range.

\paragraph{Why two methods at once, and which one is better?} The answer to the second question: it depends. A good deal of the time processes is divided into two large classes.
\paragraph{AR-processes} are characterized by the point that the current value of the process is determined by one or more past values closest to it.

Imagine a football team whose fans are far from fanaticism and support it more or less ardently, depending on the joy the team brings to them. Let's assume that the team has played fine recently, so that the last three games collected a large number of fans in the stands. And suddenly... the next two games were unsuccessful. But even the second of them has collected quite a lot of people (though some less then the previous great games). The thing is that many fans have come maintaining euphoria from those three wonderful games in the past. They hoped that the latest failures were accidental, and a new game would be so great as those three. In terms of (\ref{eq:AR}), if \textit{q}=3, the number of fans ${X}_t$ attending a new game  must be big, since the three terms on the right-hand side are big. If the losing streak continues, the number of fans will steadily decrease even if luck will come back to the team.

\paragraph{MA-processes.} To comprehend them, imagine a bit differently. Let the fans of our team be really ardent fans, and the desire to visit and support the team be always great, regardless of how it plays. But any fanaticism has its limits: even the coolest guy won’t look forward to get sick after sitting  in the stands under rain, hail, and snow at the same time... Let's assume that we are having a warm, dry, and nice April. The weather service has always been giving forecasts exactly in this vein.

And suddenly --- with an excellent weather forecast for the day of another game --- as soon as the game began --- the stadium got covered with rain, hail, snow and tornadoes... Thus, the weather service's error provoked some fans to make their own \textit{errors} refraining from attending the \textit{next} game, despite their fanaticism. So the MA model contains an error made at the current moment ($\epsilon_t$) based on the erroneous forecast for this moment made the day before ($\epsilon_{t-1}$).

Of course, the non-attendance of some fans at the new game may be partly due to the fact that they could have been `burned' by nasty weather previously too (values $\epsilon_{t-2} \div \epsilon_{t-q} $).

\subsection{Autocorrelation, Partial Autocorrelation}

\paragraph{So how many terms should expressions (\ref{eq:AR})-(\ref{eq:MA}) contain?}
Suppose, the MA model has them four \smiley:

\begin{gather}
	{X}_t=\mu+\epsilon_t+\theta_1 \epsilon_{t-1}+\theta_2 \epsilon_{t-2}\label{ma_0}, \\
	{X}_{t-1}=\mu+\epsilon_{t-1}+\theta_1 \epsilon_{t-2}+\theta_2 \epsilon_{t-3}, \\
	{X}_{t-2}=\mu+\epsilon_{t-2}+\theta_1 \epsilon_{t-3}+\theta_2 \epsilon_{t-4}, \\
	{X}_{t-3}=\mu+\epsilon_{t-3}+\theta_1 \epsilon_{t-4}+\theta_2 \epsilon_{t-5}\label{ma_3},
\end{gather}
and ${X}_{t}$ correlates with ${X}_{t-1}, {X}_{t-2}$ since their expressions include common components: errors at the same moments (respectively, $\epsilon_{t-1}$, $\epsilon_{t-2}$;\:\: $\epsilon_{t-2}$).

The same cannot be said about the pair ${X}_{t} - {X}_{t-3}$ because they have no common components. The errors at different time points, being random, have zero correlation. Thus, \textit {autocorrelation}, i.e., the correlation of a series with itself shifted by a certain lag, equals zero for all the lags 3 and greater.

The Python module \textit{statsmodels} offers tools for visualizing time series autocorrelation functions \textit{ACF} and \textit{PACF} relating the autocorrelation value of the series with the lag:
\begin{lstlisting}[language=Python]
from statsmodels.graphics.tsaplots import plot_acf, plot_pacf
#...
\end{lstlisting}
From the above said it follows that the length of the series (\ref{eq:AR}), or \textit{the model order}, should be chosen as the maximum lag at which the autocorrelation is still statistically significant, i. e., located outside the confidence interval. In Fig. \ref{pic:acf_init}, for example, we would choose 7 (see also \cite{Zvornicanin}).

Expressions similar to (\ref {ma_0})--(\ref {ma_3}) can also be written for the AR model (this time we conditionally set the order equal to 3):

\begin{gather*}
	{X}_t=\phi_0+\phi_1 X_{t-1}+\phi_2 X_{t-2}+\phi_q X_{t-3}, \\
	{X}_{t-1}=\phi_0+\phi_1 X_{t-2}+\phi_2 X_{t-3}+\phi_3 X_{t-4}, \\
	{X}_{t-2}=\phi_0+\phi_1 X_{t-3}+\phi_2 X_{t-4}+\phi_3 X_{t-5}, \\
	{X}_{t-3}=\phi_0+\phi_1 X_{t-4}+\phi_2 X_{t-5}+\phi_3 X_{t-6}, \\
	{X}_{t-4}=\phi_0+\phi_1 X_{t-5}+\phi_2 X_{t-6}+\phi_3 X_{t-7}.
\end{gather*}
Note that  expressions for pair $ {X} _ {t} - {X} _ {t -4} $ do not contain common components. Does this mean that the correlation between them is zero as in the case of the MA model? By no means: ${X} _ {t-3}$, according to the law of the model, can be expressed through $ {X} _ {t-4}, X_ {t-5}, X_ {t-6} $. So $ {X} _ {t} $ does depend on them. It is easy to see that formally autocorrelation is nonzero with a whatever large lag.

Therefore, the order of the AR model, similarly to the MA case,  is detected using the \textit{PACF (partial autocorrelation)} function. It gives an autocorrection for a certain lag, from which all the \textit{autocorrelations for the intermediate smaller lags} are excluded. Fig. \ref{pic:pacf_init}, apparently, gives order 1.

\begin{figure}[!h]
	\centering
	\begin{subfigure}[]{0.48\textwidth}
		\includegraphics[width=1\textwidth]{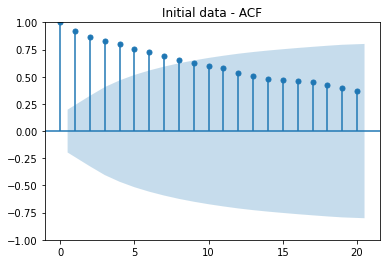 }
		\caption{ACF of original time series}
		\label{pic:acf_init}
	\end{subfigure}
	\begin{subfigure}[]{0.48\textwidth}
		\includegraphics[width=1\textwidth]{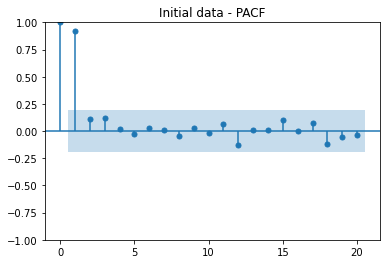 }
		\caption{PACF of original time series}
		\label{pic:pacf_init}
	\end{subfigure}

	\begin{subfigure}[]{0.48\textwidth}
		\includegraphics[width=1\textwidth]{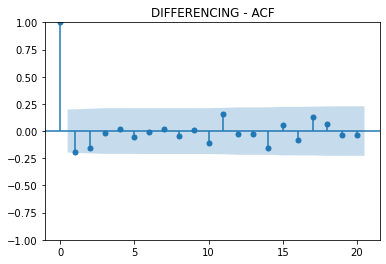}
		\caption{ACF of differentiated time series}
		\label{pic:acf_diff}
	\end{subfigure}
	\begin{subfigure}[]{0.48\textwidth}
		\includegraphics[width=1\textwidth]{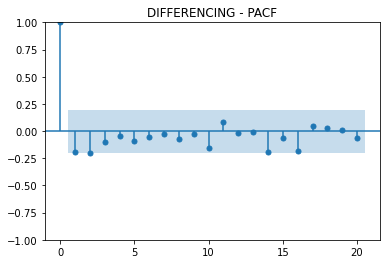 }
		\caption{PACF of differentiated time series}
		\label{pic:pacf_diff}
	\end{subfigure}
	\caption{Autocorrelation functions in action.}
	\label{fig:augmentation}
\end{figure}

Look at Fig.~\ref{pic:acf_diff}, \ref{pic:pacf_diff} and appreciate diminishing \textit{ACF} and \textit{PACF} \textit{after differencing} our original series. Not only does this effective procedure make a series stationary, but also supplies models with smaller orders. That makes them simpler and more stable while fitting. In our case, \textit{PACF} is so small that Fig.~\ref{pic:pacf_diff} gives 0 for the order of the differentiated time series AR model. What will the AR model be like in this case? It will not be at all! We have a purely MA process in this case. Most real processes, though, combine AR and MA aspects. For their description, more complex models, integrating the above discussed AR and MA, are used.

Before proceeding to them, pay attention to the fact that the name `moving average' belongs to both the model (\ref{eq:MA}) and the procedure of making a time series stationary (section \ref{par:ma}). What is common between them? It turns out that nothing. \cite{Jenkins, Pankratz} point out that MA is a technically incorrect name for the statistical model  (\ref{eq:MA}) caused by some obscure historical reasons.

Perhaps it seemed to someone in the past that, that if the mean of the series  $\mu$ is 0, then the sum of the rest of the terms is similar to a weighted average of the error over the time range $t-q\;\;\div\;\; t$. However, the `weights' $\theta$ are not obliged to be positive or to give 1 in sum. Therefore,  generally, this expression is not a weighted average ...

\subsection{Integrated statistical methods}
\subsubsection{ARMA}
\noindent is used for processes that have AR and MA aspects. The model simply  combines additively AR and MA, i.e., sums up expressions (\ref{eq:AR}) and (\ref{eq:MA}). Thus has two parameters: AR and MA orders $ARMA(p,q)$.

\subsubsection{ARIMA (Autoregressive Integrated Moving Average)}
\label{sec:arima}
We talked about meaning and profits of differencing in section~\ref{sec:diff}. Like ARMA, ARIMAl integrates AR, MA, adding the third parameter \textit{d}, indicating how many times the series needs to be differentiated to achieve stationarity.

Python libraries have a powerful type \textbf{$pmdarima$.$auto\_arima$} constructing the best model (based on the `synthetic' Akaike Information Criterion) using \textit{\textbf{grid search}}, i.e., trying out all possible combinations of the model orders within reasonably specified ranges.

\subsubsection{SARIMA (Seasonal Autoregressive Integrated Moving Average)}
\noindent  is intended to take into account seasonality combining multiplicatively two ARIMA models. The first one describes the trend, and the second is for seasonal components of the time series. (Fig. \ref{fig:decomp_init}):
\begin{equation*}
ARIMA (p, d, q)(P, D, Q)_m\ ,
\end{equation*}

\begin{itemize}
	\item[\ding{226}] \textit{p} --- non-seasonal AR order,
	\item[\ding{226}]  \textit{d} --- non-seasonal MA order,
	\item[\ding{226}]  \textit{q} --- non-seasonal differencing order,
	\item[\ding{227}]  \textit{(P, D, Q)} --- correspondent seasonal orders,
		\item[\ding{165}]  \textit{m} ---  the number of time steps in a seasonal cycle; typically can take values 7 (Daily), 	12 (Monthly), 52 (Weekly), 4 (Quarterly), 1 (Annual, non-seasonal).
\end{itemize}

\subsection{Results}
	\label{sec:results}

The best results were obtained for the series with aggregation periods of 10 and 12 months using ARIMA with `manual' selection of parameters based on ACF-PACF graphs. The parameters  suggested by the $auto\_arima$ object as the best ones,  gave worse forecasts.

For the 6 months interval, the forecast was only able to guess the test data mean. Despite the fact that the model fitted quite accurately for  the arcsined and restored data  (Fig. 	\ref{fig:init6}, 	\ref{fig:arc6}).

\begin{figure}[!h]
	\centering

	\begin{subfigure}[]{0.32\textwidth}
		\includegraphics[width=1\textwidth]{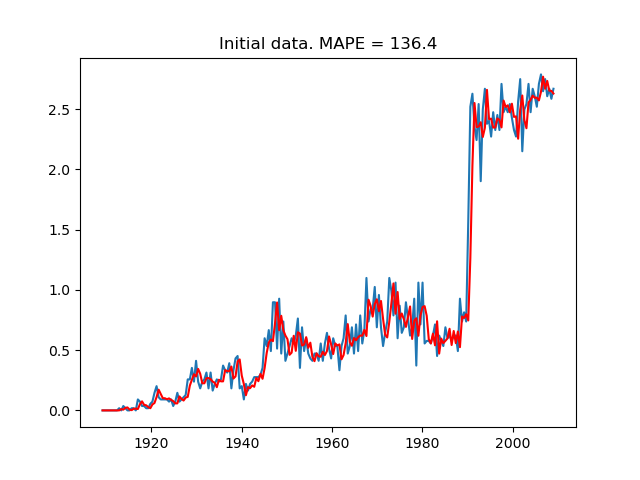 }
		\caption{Training of\\ $\arcsin$-transformed series}
		\label{fig:arc6}
	\end{subfigure}
	\begin{subfigure}[]{0.32\textwidth}
		\includegraphics[width=1\textwidth]{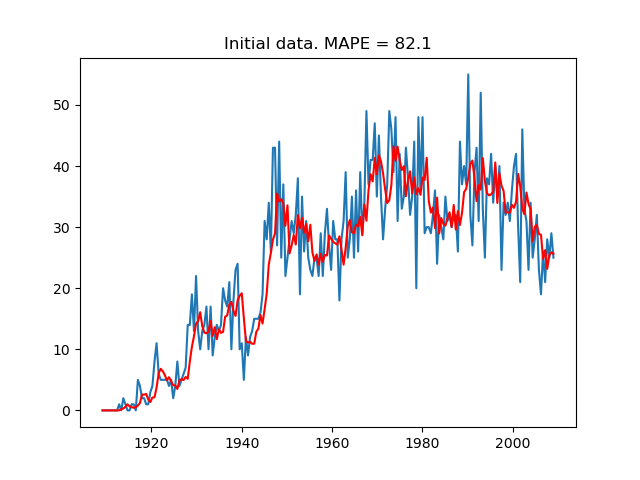 }
		\caption{Training of \\restored series}
		\label{fig:init6}
	\end{subfigure}
		\begin{subfigure}[]{0.32\textwidth}
		\includegraphics[width=1\textwidth]{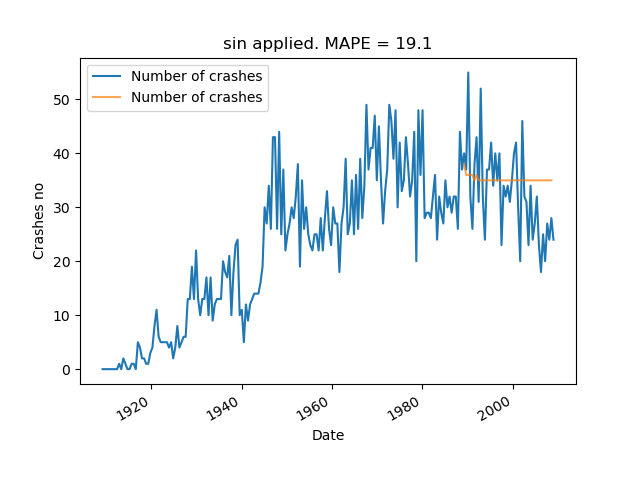 }
		\caption{Forecast}
	\end{subfigure}
	\caption{ARIMA test forecast. Interval of 6 months, \\test data portion 20\% }
	\label{fig: arima_res_6}
\end{figure}

\begin{figure}[!h]
	\begin{subfigure}[]{0.49\textwidth}
		\includegraphics[width=1\textwidth]{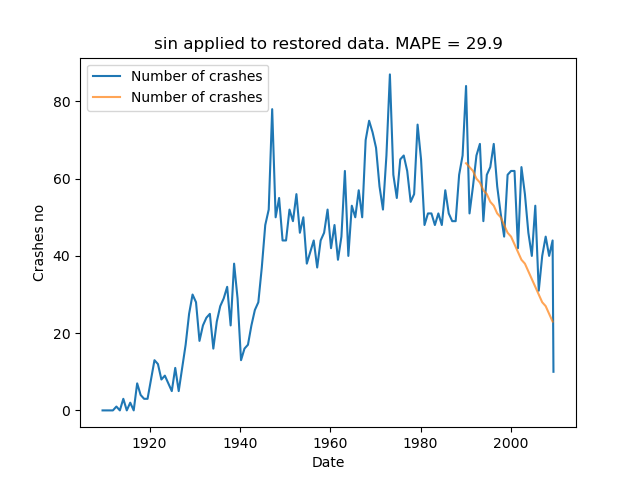 }
		\caption{Interval of 10 months, test data \\portion 20\% }
	\end{subfigure}
	\begin{subfigure}[]{0.49\textwidth}
		\includegraphics[width=1\textwidth]{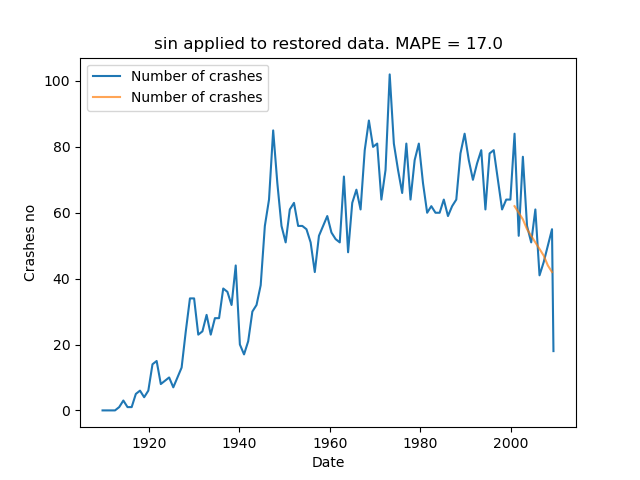  }
		\caption{Interval of 12 months, test data \\portion  10\% }
	\end{subfigure}
		\caption{Better test predictions of the ARIMA model}
		\label{fig: arima_res_10_12}
\end{figure}

For intervals of 10 and 12 months, the forecast captures a trend that is not explicitly seen in the historical data: a sharp decrease in the number of disasters.

Yet, the results are not amazing. But go on...

\section{Time forecast by means of deep learning}
\subsection{The most known nowadays types of recurrent neural networks }
\subsubsection{Simple RNN  }

Even in the simplest version, a recurrent network can be represented as a `deck' of networks, each being trained with data from its own moment in time, taking into account the previous moment network training  (Fig. \ref{fig: rnn}). The learning result is transmitted to the current network using the so-called \textit{Delay Units} \cite{Lewis}.

\begin{figure}[h]
	\centering
	\includegraphics[width=0.6\textwidth]{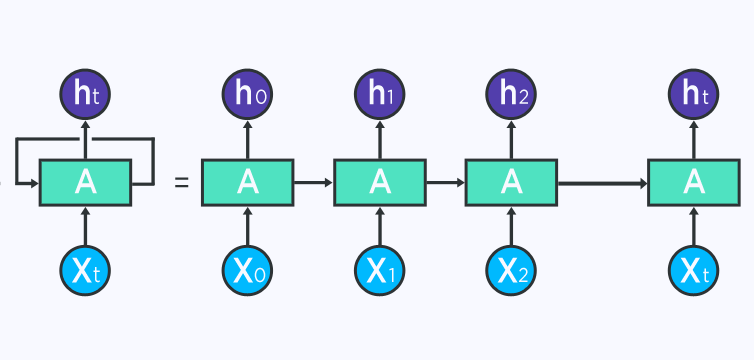 }\\
	\caption{Expanded representation of a recurrent neural network, according to \cite{Donges}}
	\label{fig: rnn}
\end{figure}

The simple RNN has thus a short memory. \textit{Directly,} it remembers only what happened a period ago, although \textit{indirectly}, of course the rest of the past is kept too.

\subsubsection{LSTM (long short-term memory) networks}
\noindent like RNN, have a short memory (output of the hidden layer $h_{t-1}$, Fig. \ref{fig: lstm}). But there is a long one as well! The Cell State, depicted on top of the figure by a horizontal line, is like a brain moving along a conveyor belt and undergoing at each stop \textit{learning} the news of the current moment (partially or completely) and \textit{forgetting} elements of the past (partially or completely). The LSTM architecture suggests that this brain will end up with the most adequate memory of the total past.

\begin{figure}[h]
	\centering
	\includegraphics[width=0.6\textwidth]{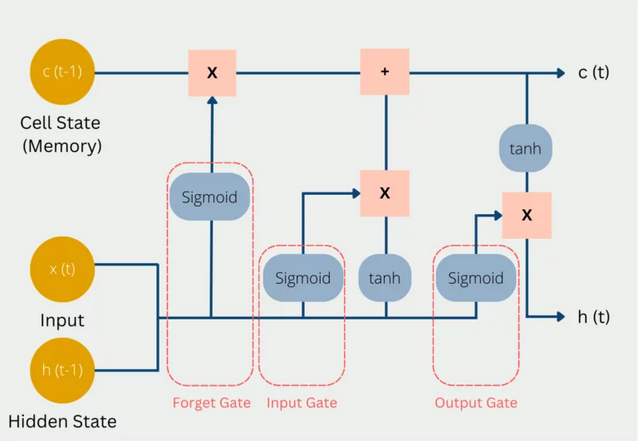 }\\
	\caption{ LSTM architecture, on \cite{lstm}}
	\label{fig: lstm}
\end{figure}

What to add to the long-term memory and what to remove from it --- and to what extent --- is regulated by structures called gates. Consider their work step by step.

\paragraph{The first step} of the LSTM algorithm is to decide what information from the long memory state of the previous moment we are going to throw away now. This decision is made by a $sigmoid$-activated gate called \textit{\textbf{forget gate}}. On the basis of  $h_{t-1}$ and the input value $x_t$, for each memory component of $C_{t-1}$, it outputs the result ranging from 0 (get rid of this element completely) to 1 (keep this element completely). Restate this in formal terms using the notation traditional in the literature on neural networks:

\begin{equation*}
	f_t=\sigma(W_f\cdot [h_{t-1},x_t]+b_f).
	\label{eq:forget}
\end{equation*}

\paragraph{The next step} is to determine what information we are going to update in the long memory. To begin with, the sigmoid layer called \textit{\textbf{input gate}} decides which values and to what extend should be updated. Then, the layer with the $\tanh$-activation creates a vector of new values-\textit{candidates} $\tilde{ C } _t$ to be added to the memory (it will be corrected in the following step):
\begin{gather*}
	i_t=\sigma(W_i\cdot [h_{t-1},x_t]+b_i), \\
    \tilde{ C } _t= \tanh (W_C\cdot [h_{t-1},x_t]+b_C).
\end{gather*}

\paragraph{Finally,} refresh the memory state. The new state $C_{t}$ will be a weighted sum of the old state and the candidate for the new state. We multiply the old state by $f_t$, forgetting what we have decided to forget earlier (in the needed measure). Do similarly with the candidate for the new state using $i_t$:

\begin{gather*}
C _t= f_t * C _{t-1} +i_t * \tilde{ C }_t.
\end{gather*}

\noindent In the long run, we will care about the short memory, based on the new state $C_t$.  Using here the \textit{\textbf{output gate}} which will decide what elements of the new long memory state are more or less relevant for the \textit{nearest} future. Then form a new short memory:

\begin{gather*}
o_t=\sigma(W_o\cdot [h_{t-1},x_t]+b_o),\\
h_t=o_t*\tanh(C_t).
\end{gather*}

\subsubsection{Stateful LSTM networks}

Of interest is the ability of LSTM to accumulate and maintain its state between training on different pieces of data, \textit{the batches}, so that the network learns on the current batch `keeping in mind' the previous batches. There is a nuance in their use in Python: \textit{batch sizes} must be the same when compiling the network and when using it. \cite{Brownlee} explains how to properly meet this nuance.

With our task, unfortunately, stateful LSTM networks failed to perform brilliantly.

\paragraph{Bidirectional LSTM networks}
just combine pairs of LSTM networks learning in opposite directions \cite{Maheshmj}. Is it supposed herewith that the future influences the past? At any rate, the form of more historical data may well correlate with the form of modern data. This possibility is taken into account in such an architecture.

In our case, this type of networks performed most successfully (see section~\ref{sec:deep_results}).

\subsubsection{GRU (gated recurrent unit) networks}

This is a new generation of recurrent neural networks, very similar to LSTM. The GRU networks got rid of the long-term memory as an architectural element. To distinguish long-lived dependencies in a time series, the internal $h_{t-1}$ layer is used. There are only two gates, \textit{\textbf{update gate}} and \textit{\textbf{reset gate}}. They perform work similar to the LSTM gateways \textit{\textbf{forget gate}}~---~\textit{\textbf{input gate}} and \textit{\textbf{output gate}}, respectively. The result of their work is used by the internal layer $h_{t+1}$.

Fewer gates --- fewer tensor operations, which makes GRU slightly faster than LSTM. In practice, neither network is clearly superior, so it is common to try both options to see which one is better suited for a particular case. In our case, LSTMs --- uni- and bidirectional --- turned out definitely better (see discussion of the results). So let's wish GRU success in future tasks.

\subsection{Possibilities of deep learning parameterization}
\paragraph{Cross-validation for time series,}

as in standard machine learning, helps deal with overfitting. However, for time series processing, successive training sets, or folds, are formed not on a random basis, but as `supersets' of previous ones That is, each subsequent set is the previous fold plus one or more additional values (their number is determined by a combination of \textit{gap} and \textit{n\_splits} arguments):
\begin{lstlisting}[language=Python]
from sklearn.model_selection import TimeSeriesSplit
TimeSeriesSplit(gap=0, max_train_size=None, n_splits=5, test_size=None)
#...
\end{lstlisting}

\paragraph{Using multiple time steps} to predict the next step. An example can be found in \cite{Lewis}.  In  \textbf{\textit{the Code} } this option is regulated by the variable
$more\_time\_steps$  of the $crashes\_deep$.$do\_predict$ function.

\paragraph{Various activation functions.} The file \textit{crashes\_deep.py} contains 15 definitions of the functions that were used in the research. The unsaturated functions \textit{relu} and \textit{softplus}  turned out to do well as activations of the hidden layers --- much better than the default sigmoid.

\paragraph{Different distributions of initial values of weights and biases.}
Keras provides a number of ways to initialize weights and biases. Most of them are modifications of normal and uniform distributions (for example, a normal distribution with `the tails cut'). The initializations used in \textbf{\textit{the Code}} are defined in the file \textit{crashes\_deep.py}.

\subsection{Implementation and results}
\label{sec:deep_results}

\begin{figure}[h]
	\centering
	\includegraphics[width=0.6\textwidth]{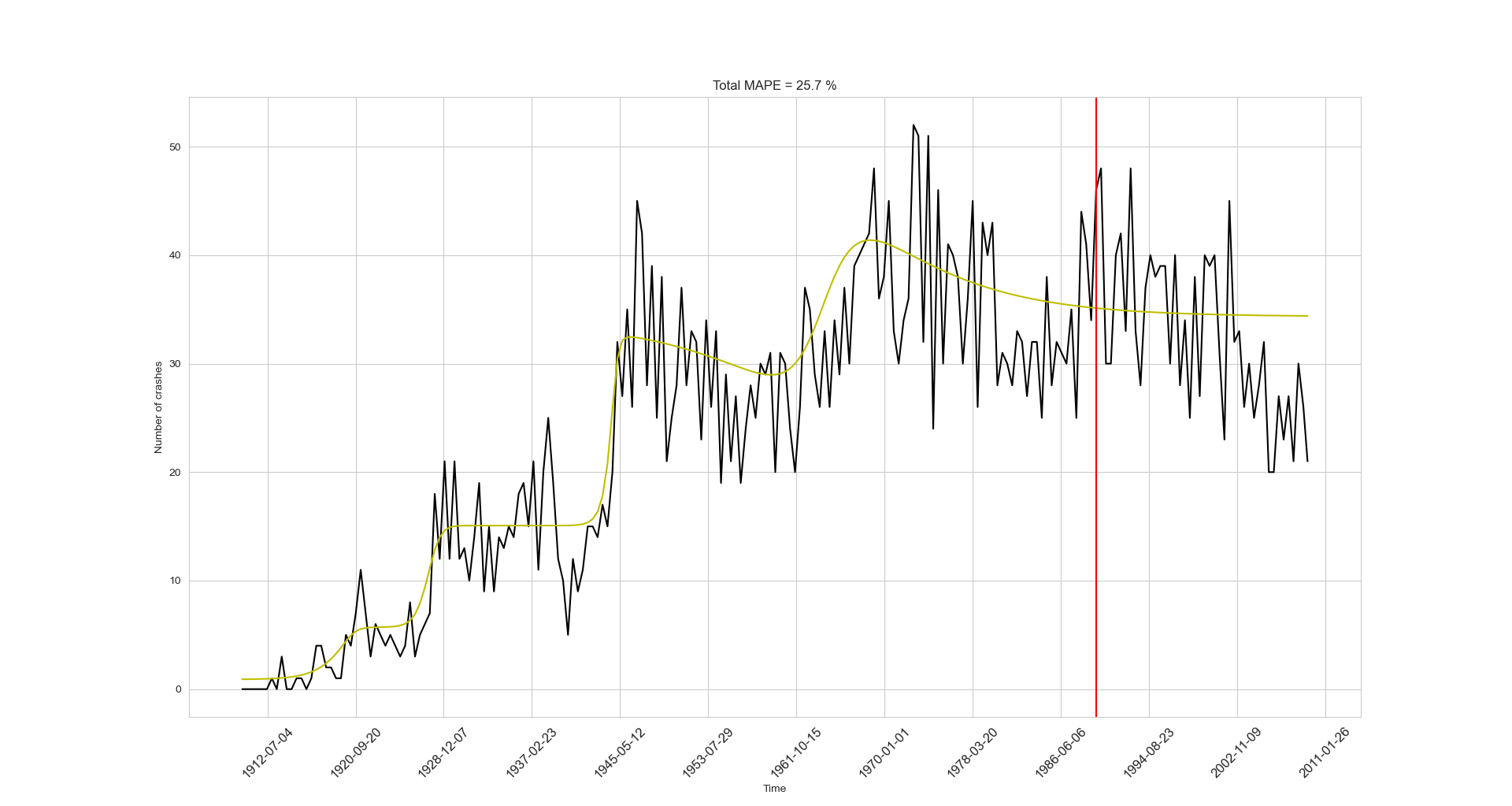  }
	\caption{GRU forecast. 6 month interval }
	\label{fig: gru}
\end{figure}

\begin{figure}[!h]
	\centering
	\begin{subfigure}[]{0.32\textwidth}
		\includegraphics[width=1\textwidth]{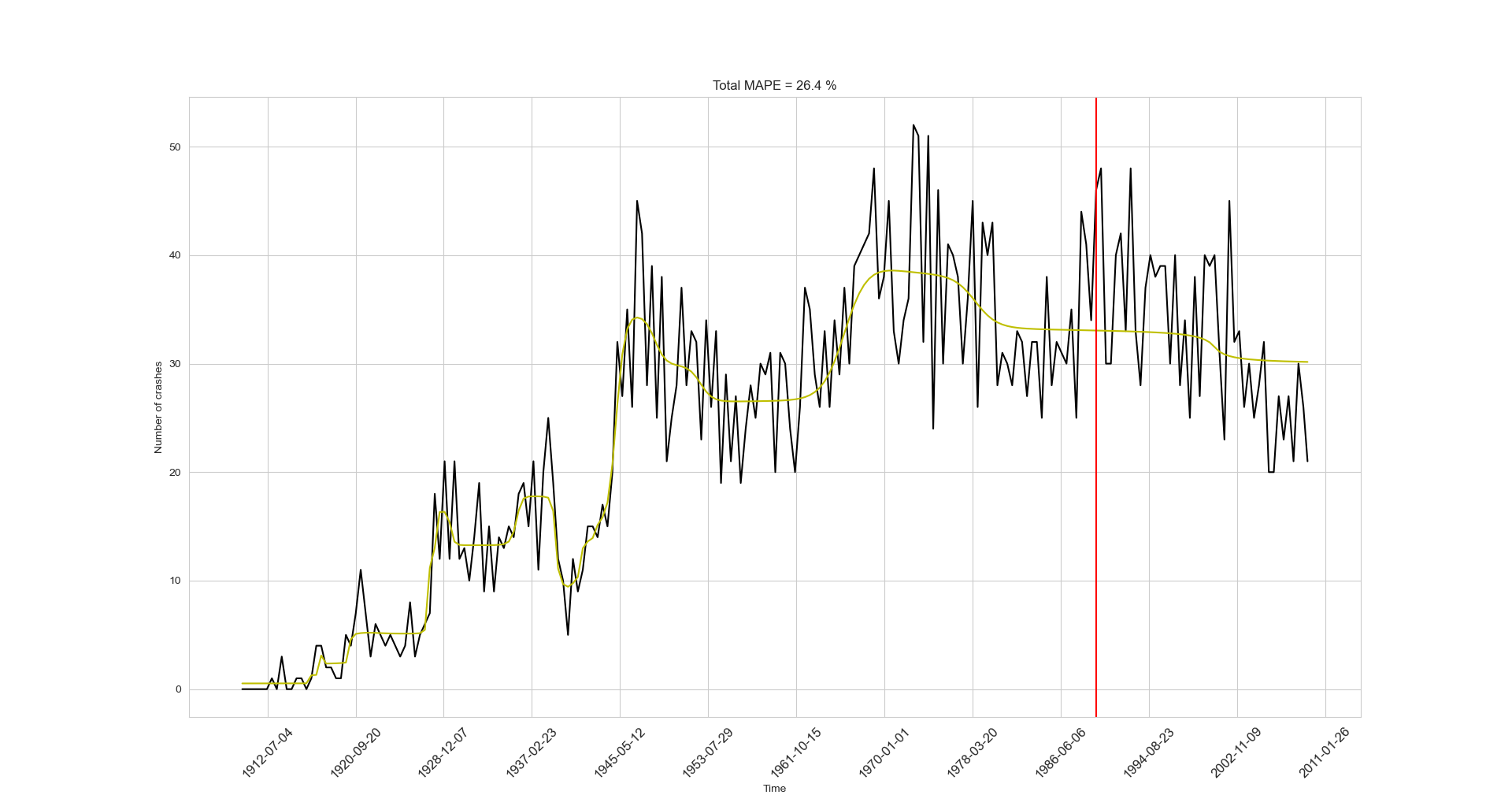 }
		\caption{ 6 month interval,\\ bidirectional LSTM}
		\label{fig:deep_res_6}
	\end{subfigure}
	\begin{subfigure}[]{0.32\textwidth}
		\includegraphics[width=1\textwidth]{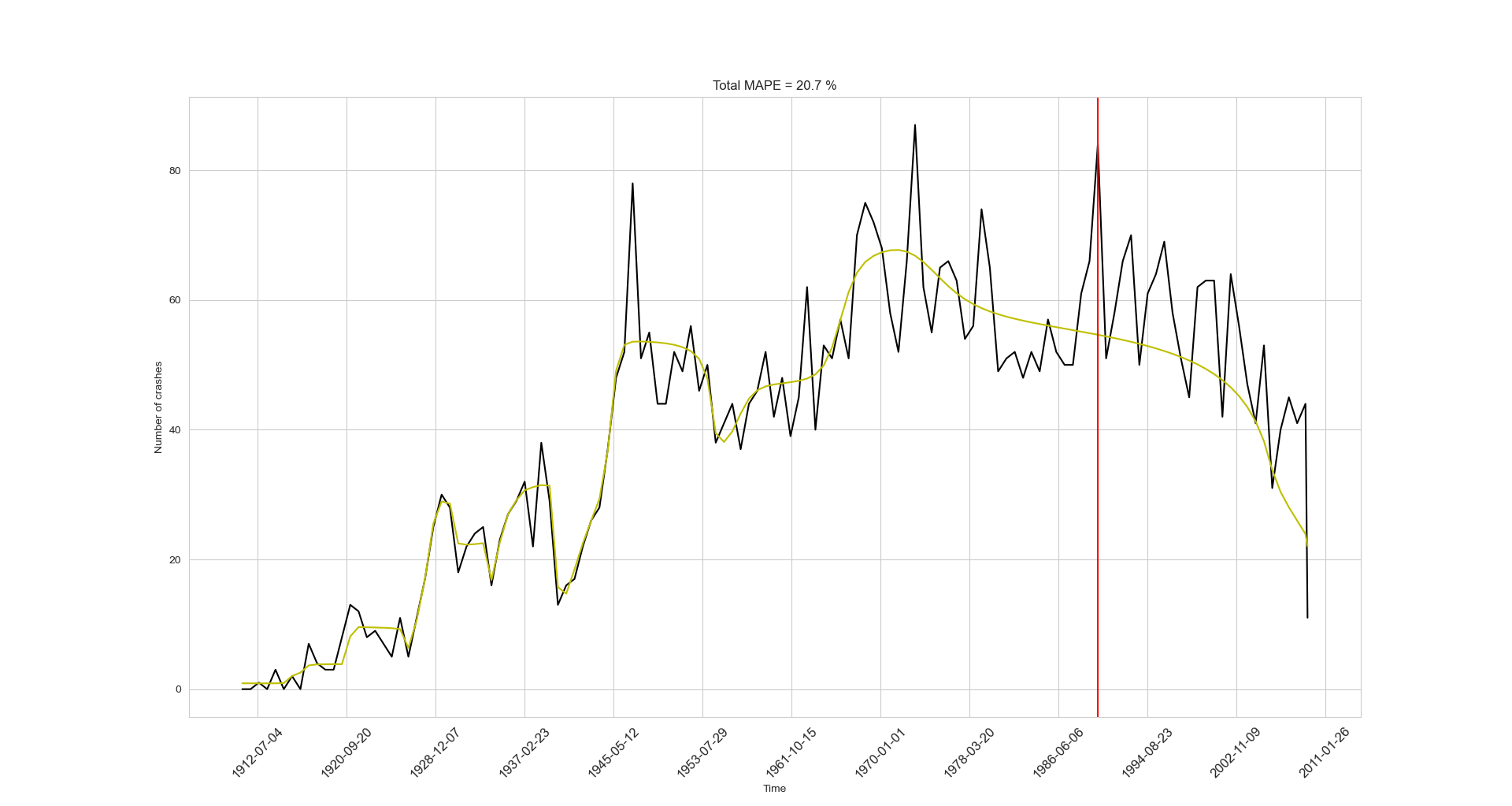}
		\caption{ 10 month interval,\\ bidirectional LSTM}
	\end{subfigure}
	\begin{subfigure}[]{0.32\textwidth}
		\includegraphics[width=1\textwidth]{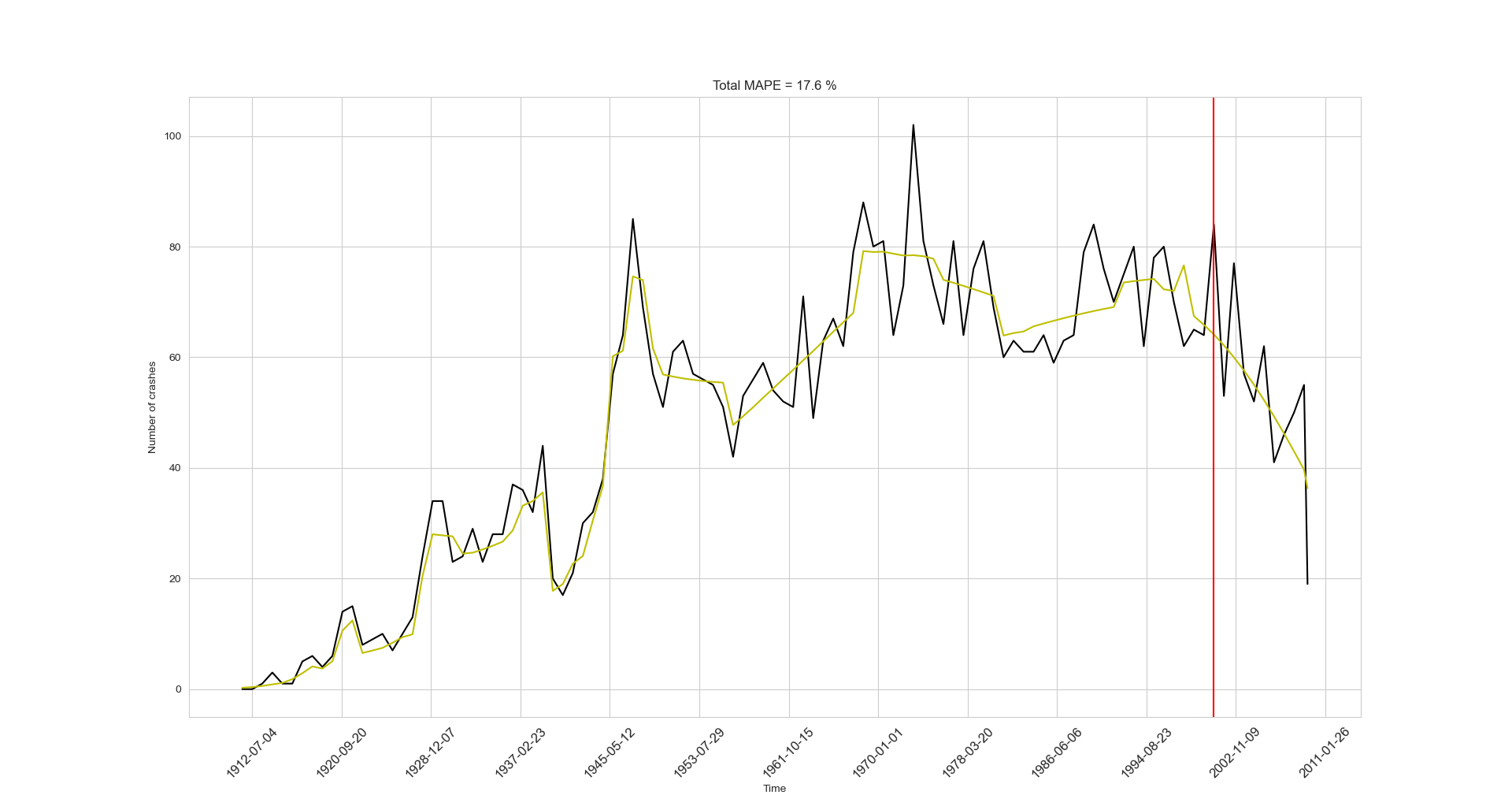 }
		\caption{12 month interval,\\ LSTM }
	\end{subfigure}
	\caption{Training and prediction of neural networks}
	\label{fig:deep_res}
\end{figure}

As noted in the discussion of GRU, these networks produced slightly worse predictions. From comparison of Fig. \ref{fig: gru} and \ref{fig:deep_res_6}, it is clear that GRU unduly roughly smooths out features of the data\footnote{To get a better look at the graphs, please zoom in as much as necessary.}. In contrast, uni- and bidirectional LSTMs correctly project the trend into the future and good average the test data (Fig. \ref{fig:deep_res}).

Still not very much of progress, in comparison with ARIMA. But let us continue.

\section{Forecast with Kernel Methods}

So, for our apparently non-linear tendency, deep learning proved a bit more successful than the linear methods.

How successful will the kernel methods \textbf{kernel ridge regression (KRR)} and \textbf{support vector regression (SVR)} be here?

Using the so-called \textit{kernel trick} \cite{Scholkopf, Theodoridis, Ranjan},  they are able, with minimal computational efforts, to provide a regression \textit{curve} of a whatever complex non-linear shape  by transfer of the problem into a hyperspace of a higher dimension (sometimes infinite-dimensional) and constructing a regression \textit{hyperplane} in it.

The main difference between the two methods is an optimization problem that is solved to find a regression hyperplane. While \textbf{KRR}, like simple linear regression, seeks to reduce the sum of squared distances from the data points to a hyperplane, \textbf{SVR} encloses a hyperplane in a so-called `$\epsilon$ channel'.
Herewith, the `$\epsilon$-insensitive' penalty function does not penalize points that are inside the channel. The rest of them are subject to a penalty proportional to the distance to the hyperplane (slack values $\xi$, $\xi^{*}$, Fig. \ref{fig: svr}). Thus, the hyperplane (and, accordingly, the hypersurface in the original space) tends to pass through the most `heaped' part of the points cloud, thereby conveying the general tendency contained in the data.

Obviously, the smaller $\epsilon$, the finer details the model is able to resolve. Fine features may contain random noise, what signifies overfitting. On the contrary, too big $\epsilon$ may result in poor accuracy. A proper trade-off is thus needed. In \textit{\textbf{the Code}} the best model is selected using \textit{\textbf{grid search}} (see section \ref{sec:arima}) to select not only $\epsilon$, but also other parameters of both KRR and SVR: \textit{ C} (regularization parameter), \textit{kernel} (kernel type), $\gamma$ (scale parameter in the kernel function), etc. (see the file \textit{kernel\_methods.py}).

\begin{figure}[h]
	\centering
	\includegraphics[width=0.6\textwidth]{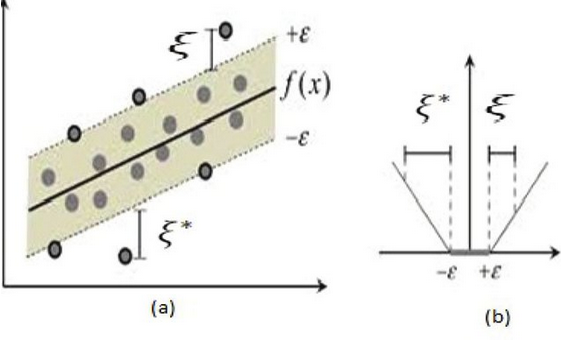}\\
	\caption{SVR: surface in the features hyperspace (a), `$\epsilon$-insensitive' \\penalty function (b). On \cite{Yasin}.    }
	\label{fig: svr}
\end{figure}

\subsection{Results and discussion}
\label{sec:kernel_results}

\begin{figure}[!h]
	\centering
	\begin{subfigure}[]{0.32\textwidth}
		\includegraphics[width=1\textwidth]{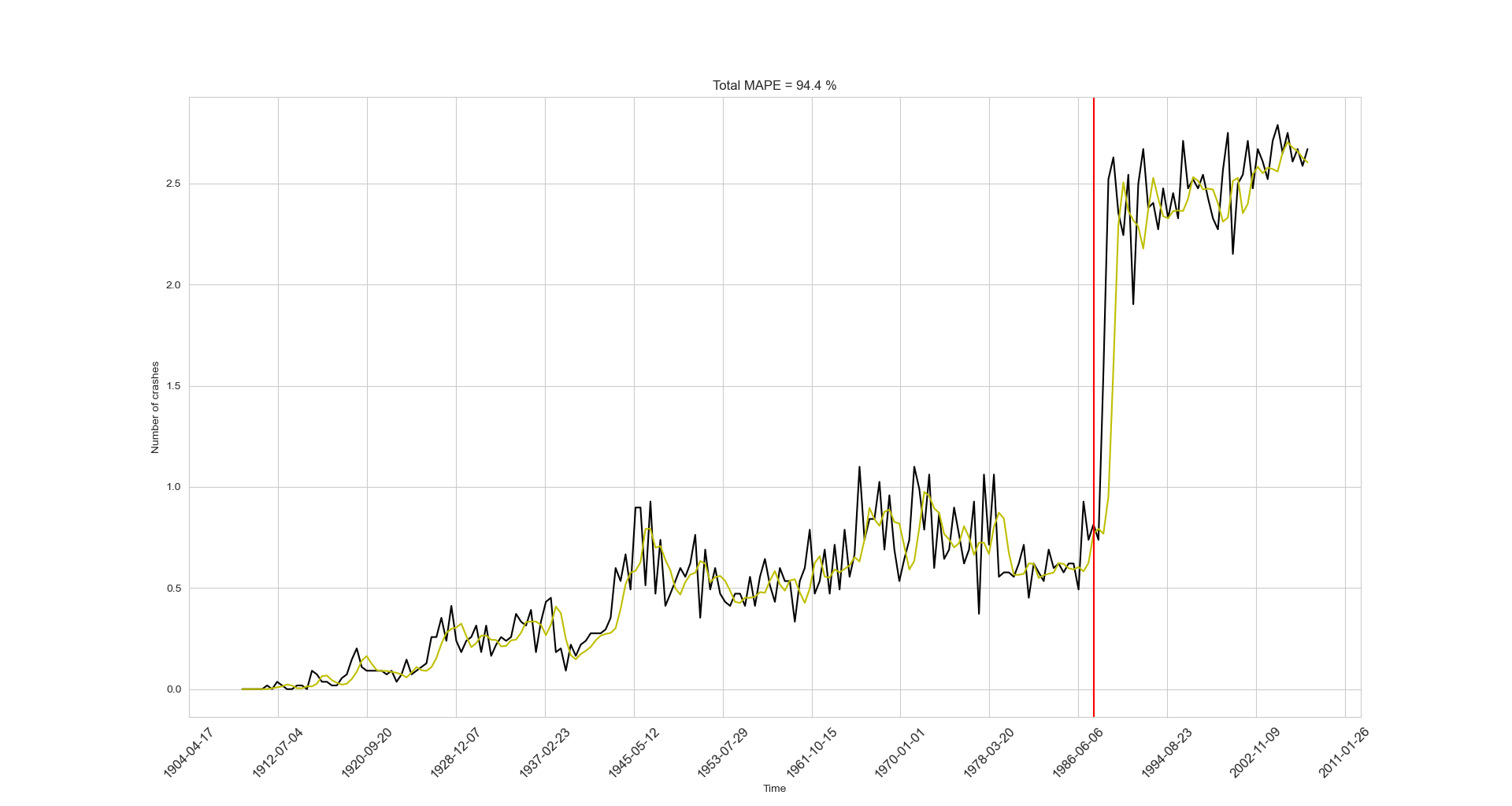 }
		\caption{6 month interval, test data portion 20\%, KRR}
	\end{subfigure}
	\begin{subfigure}[]{0.32\textwidth}
		\includegraphics[width=1\textwidth]{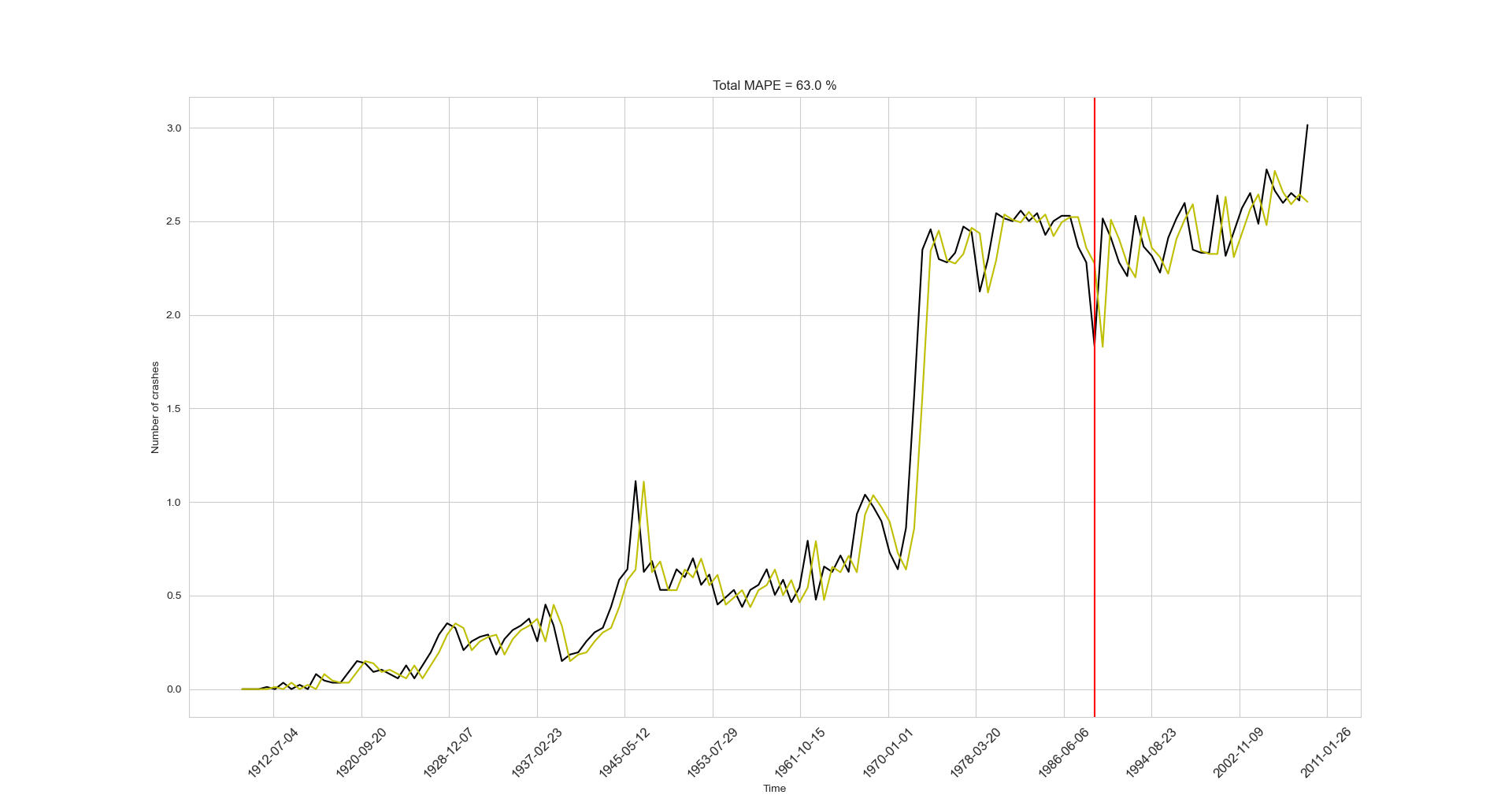}
		\caption{10 month interval, test data portion 20\%, KRR }
	\end{subfigure}
	\begin{subfigure}[]{0.32\textwidth}
		\includegraphics[width=1\textwidth]{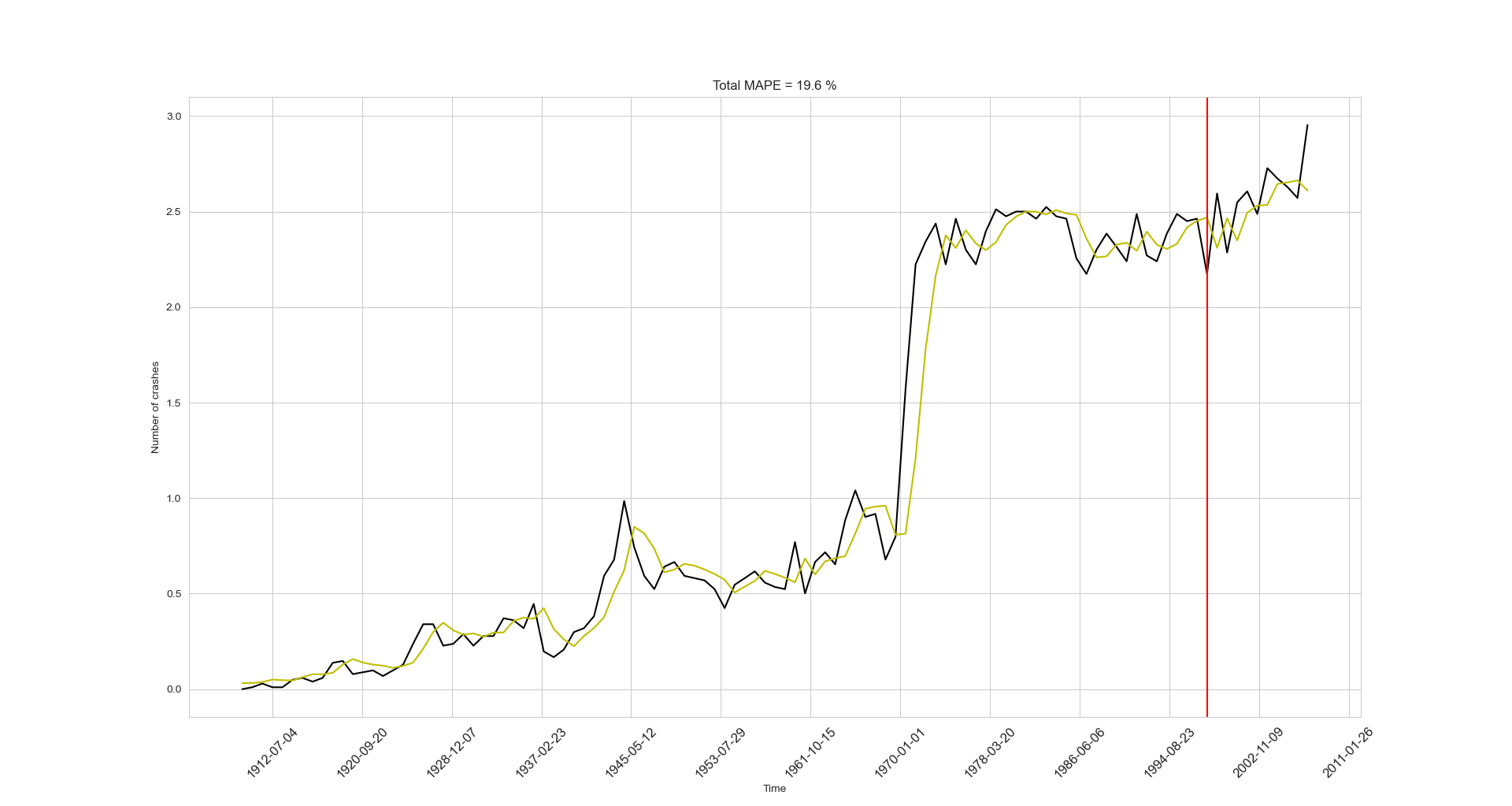 }
		\caption{12 month interval, test data portion  10\%,  SVR }
    \end{subfigure}
    	\begin{subfigure}[]{0.32\textwidth}
    	\includegraphics[width=1\textwidth]{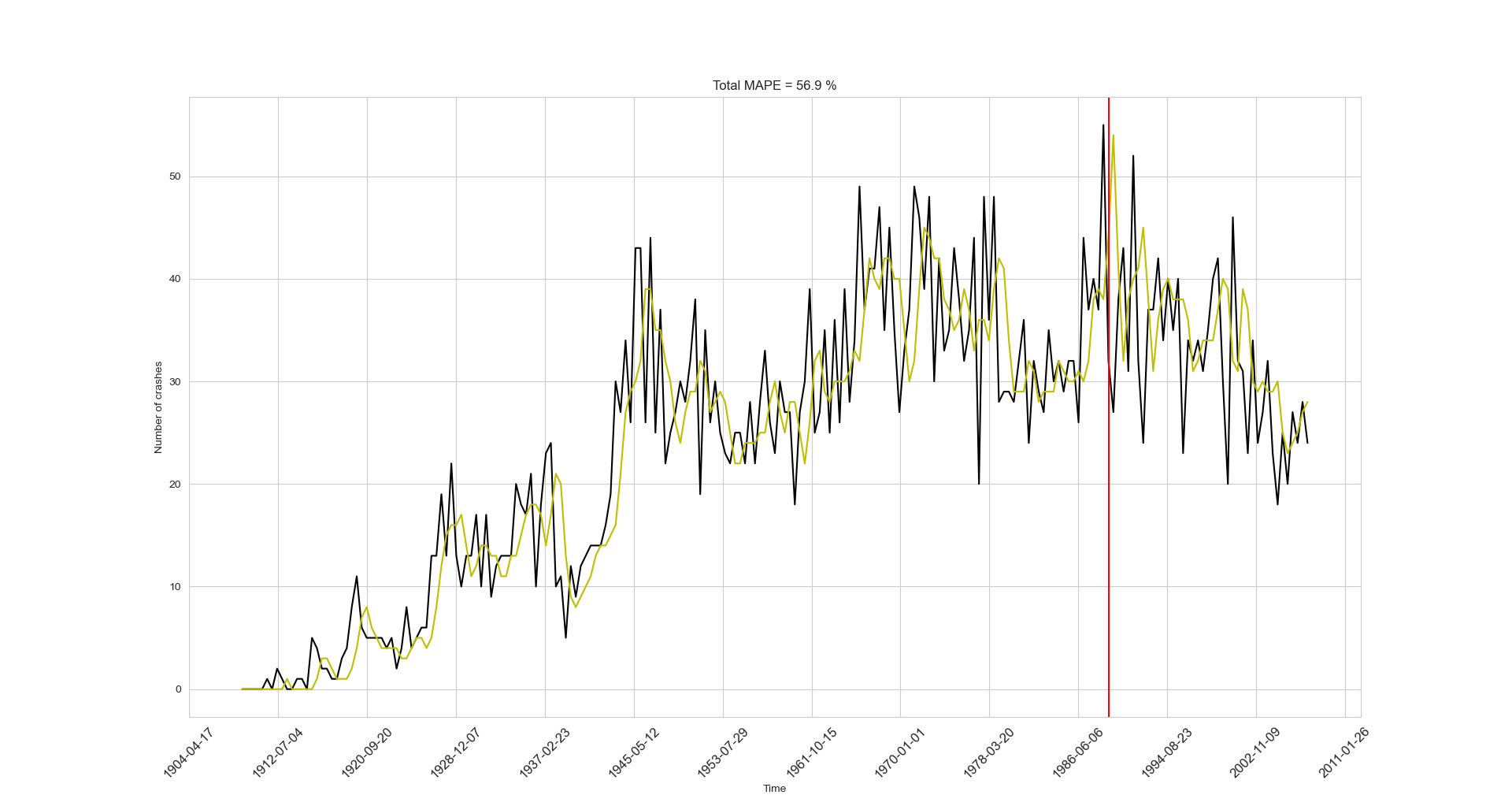 }
    	\caption{6 month interval, test data portion 20\%, KRR}
    \end{subfigure}
        	\begin{subfigure}[]{0.32\textwidth}
    	\includegraphics[width=1\textwidth]{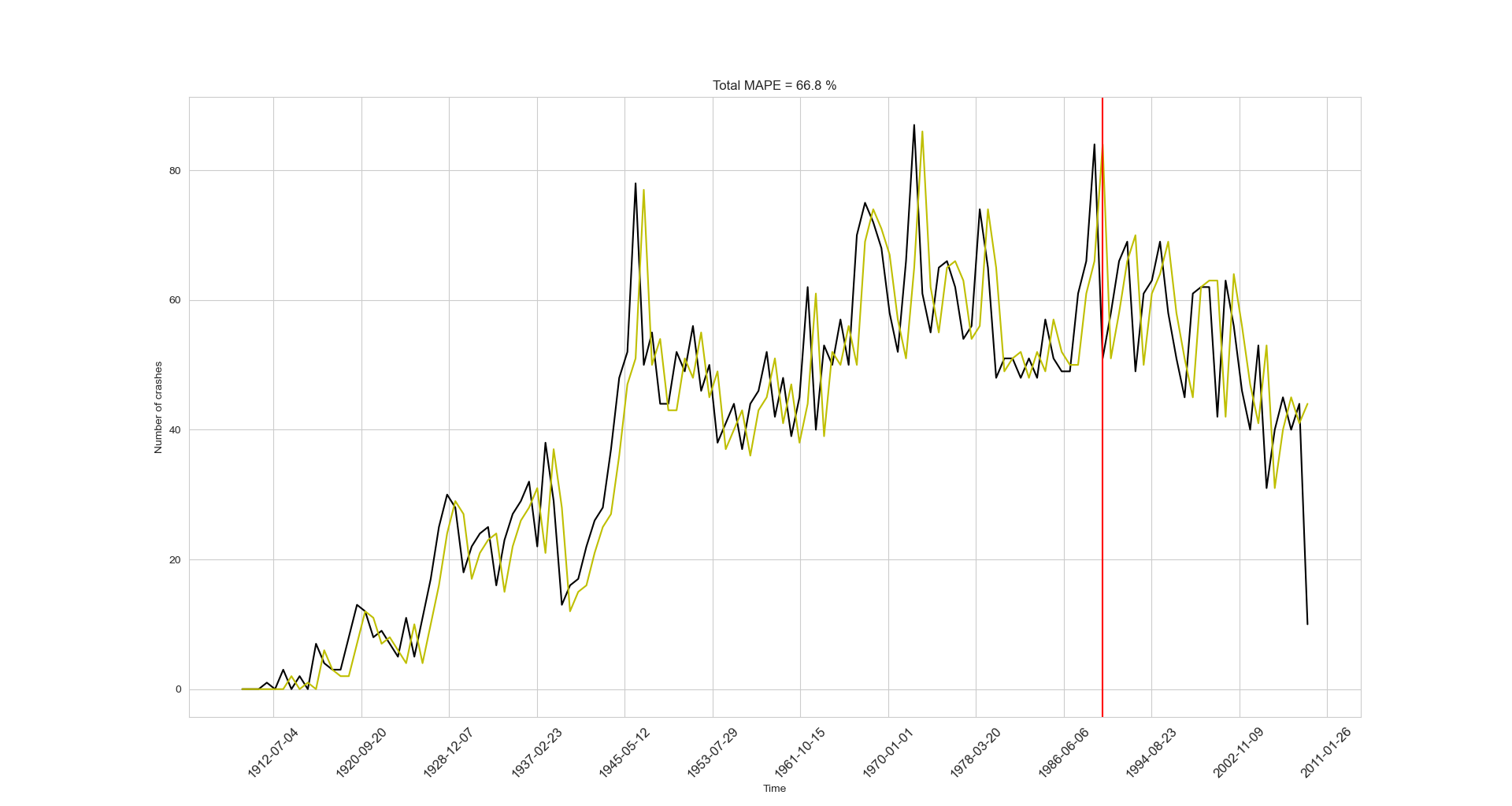 }
    	\caption{10 month interval, test data portion 20\%, KRR}
    \end{subfigure}
        	\begin{subfigure}[]{0.32\textwidth}
    	\includegraphics[width=1\textwidth]{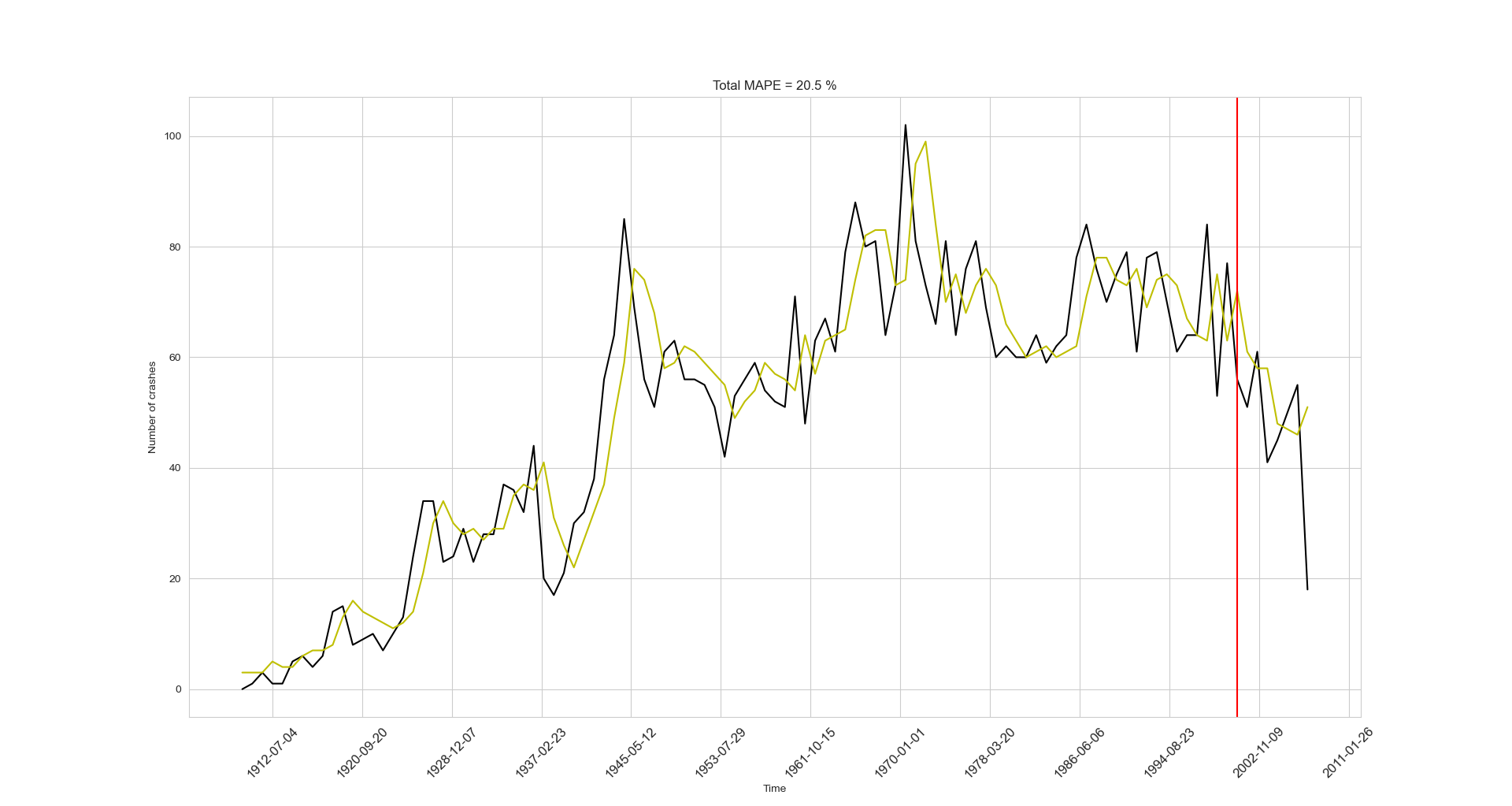 }
    	\caption{12 month interval, test data portion 10\%,  SVR }
    \end{subfigure}
	\caption{Model fitting and prediction. $\arcsin$-transformed and restored series}
	\label{fig:kern_res}
\end{figure}
The most accurate forecast was obtained when applying the $\arcsin$-transformation. The best results are shown in Fig.~\ref{fig:kern_res} for both the transformed series, for which the forecast is made, and for the $\sin$-restored data and results. As we can see, all the sharp extrema of the test data are described accurately, especially for intervals of 6 and 10 months.
\paragraph{Note.} Because of high volatility, MAPE (Mean Absolute Percentage Error) does not always duly reflect qualitative pictures of the fitting and forecast. Even a small lag between extrema of the data and forecast makes a significant contribution to the error. Perhaps a better way to quantify accuracy might be to calculate the average MAPE across some groups of time points.

\section{Conclusion}

So, we have investigated a rather sophisticated time series: quite volatile at all reasonable sampling rates. Wherein:

\begin{itemize}
	\item[\ding{224}] Detailly presented models based on AR and MA, at best, predicted satisfactorily only the general trend  of the time series test part.

	\item[\ding{224}] Neural networks, in addition to this, caught the mean  of the future values. This can be considered a good result given the specifics of the data.

	\item[\ding{224}] The kernel methods not only correctly caught the trend in the test data, but also conveyed its subtle details, albeit with a slight lag.
\end{itemize}

What has made the kernel methods so successful? Perhaps their main advantage is that they do not make difference between `older' and `newer' historical data.

This is not the case of neural networks. Connected to this is a known RNN's problem (more or less successfully solved by LSTM): the gradient vanishing \cite{Lewis}. This happens when the activation function of the hidden layer is repeatedly applied to the data while proceeding from \textit{older} to \textit{newer} time moments.

In that regard, the approach used in the design of the \textit{IgNet} neural network  \cite{Mackarov2021IgNetAS} seems promising. The gradient vanishing problem was solved there by appropriate choice of a range for initial distributions of weights and biases.

In the near future, it is planned to create a \textit{recurrent} version \textit{IgNet} and to probe it in comparison with the best existing methods for time series analysis.

To be continued.

\bibliographystyle{plain}

\bibliography{TS}
\end{document}